\documentclass[letterpaper,10pt]{article}
\usepackage{osameet2}

\usepackage{amssymb}
\usepackage{amsmath}
\usepackage{multirow}
\usepackage{upgreek}
\usepackage{gensymb}

\usepackage[pdftex,colorlinks=true,bookmarks=false,citecolor=blue,urlcolor=blue]{hyperref} 

\begin{document}

\title{Super-resolution fluorescence microscopy by stepwise optical saturation}

\author{Yide Zhang$^{1,*}$, Prakash D. Nallathamby$^{2,3,4}$, Genevieve D. Vigil$^{1}$, Aamir A. Khan$^{1}$, Devon E. Mason$^{2,5}$, Joel D. Boerckel$^{5}$, Ryan K. Roeder$^{2,3,4}$,  and Scott S. Howard$^{1,3,4,*}$}

\address{$^{1}$Department of Electrical Engineering, University of Notre Dame, Notre Dame, IN 46556, USA\\
	$^{2}$Department of Aerospace and Mechanical Engineering, University of Notre Dame, Notre Dame, IN 46556, USA\\
	$^{3}$Harper Cancer Research Institute, University of Notre Dame, Notre Dame, IN 46556, USA\\
	$^{4}$Notre Dame Center for Nanoscience and Nanotechnology (NDnano), University of Notre Dame, Notre Dame, IN 46556, USA\\
	$^{5}$Departments of Orthopaedic Surgery and Bioengineering, University of Pennsylvania, Philadelphia, PA 19104, USA\\
}

\email{$^{*}$Email: yzhang34@nd.edu, showard@nd.edu}

\begin{abstract}
	Super-resolution fluorescence microscopy is an important tool in biomedical research for its ability to discern features smaller than the diffraction limit. However, due to its difficult implementation and high cost, the universal application of super-resolution microscopy is not feasible. In this paper, we propose and demonstrate a new kind of super-resolution fluorescence microscopy that can be easily implemented and requires neither additional hardware nor complex post-processing. The microscopy is based on the principle of stepwise optical saturation (SOS), where $M$ steps of raw fluorescence images are linearly combined to generate an image with a $\sqrt{M}$-fold increase in resolution compared with conventional diffraction-limited images. For example, linearly combining (scaling and subtracting) two images obtained at regular powers extends resolution by a factor of $1.4$ beyond the diffraction limit. The resolution improvement in SOS microscopy is theoretically infinite but practically is limited by the signal-to-noise ratio. We perform simulations and experimentally demonstrate super-resolution microscopy with both one-photon (confocal) and multiphoton excitation fluorescence. We show that with the multiphoton modality, the SOS microscopy can provide super-resolution imaging deep in scattering samples.
\end{abstract}

\section{Introduction}
\label{sec:introduction}

Super-resolution fluorescence microscopy techniques, such as stimulated emission depletion (STED) microscopy \cite{Hell1994, Klar2000} and its related reversible saturable optical fluorescence transitions (RESOLFT) microscopy \cite{Hofmann2005}, photoactivated localization microscopy (PALM) \cite{Betzig2006}, stochastic optical reconstruction microscopy (STORM) \cite{Rust2006}, and structured illumination microscopy (SIM) \cite{Gustafsson2000, Gustafsson2005}, have enabled a dramatic development in modern biology by being able to discern fluorescent molecules or features that are closer together than the diffraction limit \cite{Hell2007,Balzarotti2017}.
Many super-resolution techniques, however, only work well on thin and nearly transparent samples \cite{Gigan2017}.
Super-resolution imaging deep in scattering samples is still a challenge, since optical aberrations and scattering severely degrade resolution and signal-to-noise ratio (SNR), so obtaining even diffraction-limited performance is difficult \cite{Winter2014}.
Meanwhile, the implementation of most super-resolution techniques is difficult and expensive, which hinders the universal application of super-resolution microscopy in many labs.

To increase the imaging depth, some super-resolution methods have been combined with multiphoton microscopy (MPM) \cite{Egner2002,Winter2014,Nguyen2015,Field2016}, an imaging technique widely used in biomedical research for its inherent 3D resolution, deep penetration, and minimal phototoxicity \cite{Denk1990,Zipfel2003,Hoover2013}.
On the other hand, to reduce the complexity and cost of super-resolution microscopy, novel techniques such as saturated excitation (SAX) microscopy and its variants \cite{Fujita2007,Humpolickova2009,Nguyen2015,Vigil2017,Oketani2017} and fluorescence emission difference (FED) microscopy \cite{Kuang2013} have been developed, whose implementations are easier and cheaper compared to traditional super-resolution methods.
A super-resolution technique with both deep penetration and easy implementation would be desirable and widely utilized.

In this paper, we propose and demonstrate a new kind of super-resolution fluorescence microscopy using the principle of stepwise optical saturation (SOS), which is designed to work with both confocal and MPM modalities and to permit deep penetration.
In its simplest form, two-step SOS, two conventional fluorescence images are linearly combined to produce a super-resolution image with a $\sqrt{2}$-fold increase in spatial resolution.
In general, an $M$-step microscopy uses the linear combination of $M$ raw images to generate a super-resolved SOS image, which has a $\sqrt{M}$-fold increase in spatial resolution compared to diffraction-limited images.
The improvement in resolution is theoretically infinite, but practically, it is limited by the SNR performance of the resulting SOS image.
We perform simulations and experiments with both one-photon excitation fluorescence (1PEF) and two-photon excitation fluorescence (2PEF) to validate the resolution improvement.
Similar to SAX microscopy, fluorophore saturation is used in SOS microscopy but the excitation power used to cause the saturation is relatively weak, so effects like photobleaching and photodamage are reduced.
We show that the implementation of SOS microscopy is straightforward and requires neither additional hardware nor complex post-processing, which are usually required in SAX microscopy.
With the SOS methods, it is easy to produce super-resolution images using a conventional fluorescence microscope.

\section{Principle of Stepwise Optical Saturation}
\label{sec:principle}

\begin{figure}[!t]
	\centering
	\includegraphics[width=0.7\linewidth]{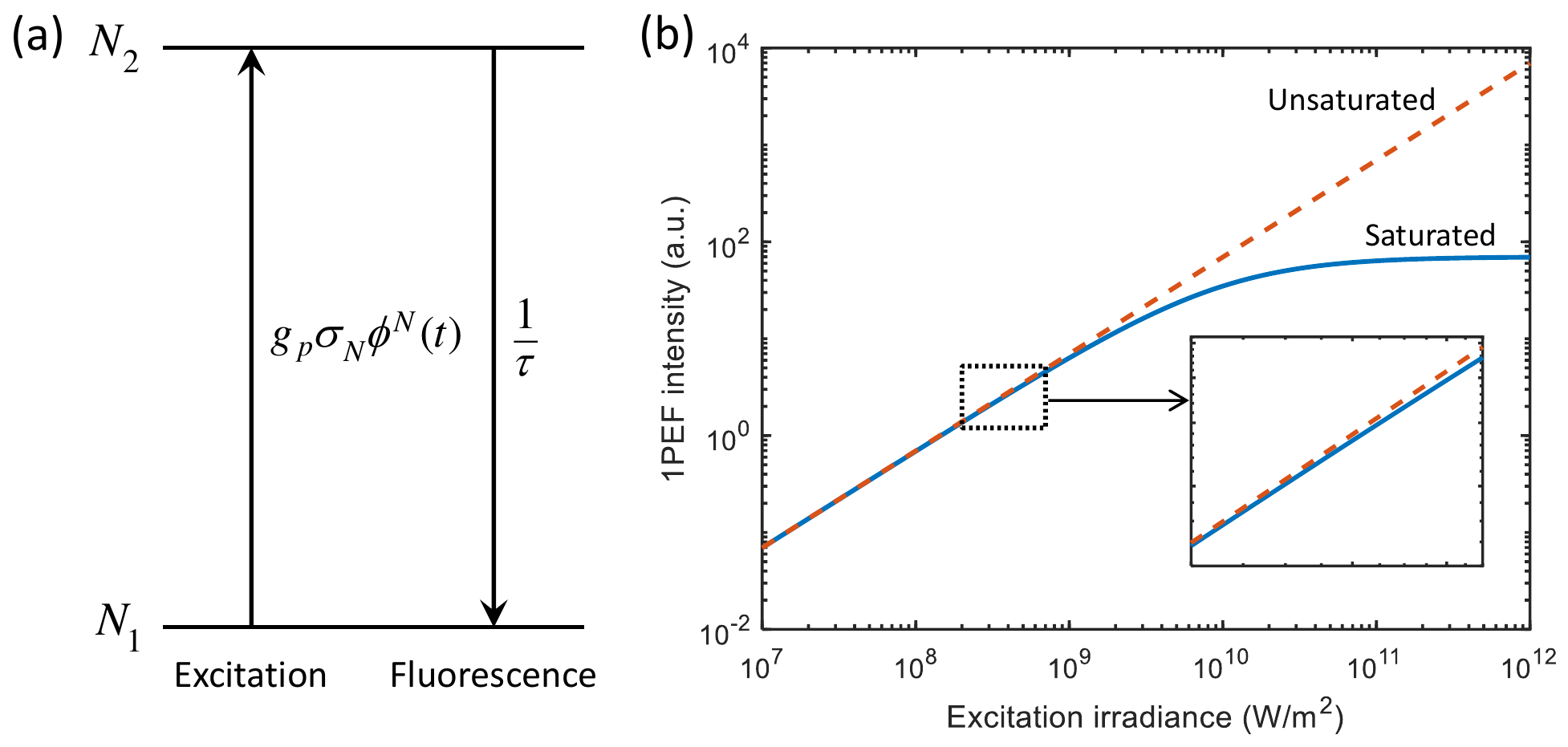}
	\caption{(a) Jablonski diagram of the two-level fluorophore model. (b) Simulated fluorescence-excitation relation for the 1PEF case showing the saturation behavior.}
	\label{fig:1_model}
\end{figure}

The principle of SOS microscopy is based on a two-level fluorophore model depicted in a Jablonski diagram in Fig. \ref{fig:1_model}(a), similar to our previous work in \cite{Zhang2017}. We denote the populations of the ground and excited singlet states as $N_1(t)$ and $N_2(t)$, respectively, as functions of time. Note that $N_1(t)+N_2(t)=N_0$, where $N_0$ is the concentration of the fluorophore. To make this model applicable to both one-photon and multiphoton excitation, the excitation rate is written as $g_p\sigma_N\phi^N(t)$, where $\phi(t)$ is the incident photon flux, $N$ is the number of excitation photons needed for a fluorophore to emit one photon ($N=1$ for 1PEF, $N=2$ for 2PEF, etc.), $\sigma_N$ is the cross-section for $N$-photon excitation, and $g_p$ is the pulse gain factor which accounts for the temporal pulse profile of the excitation \cite{Eggeling2005,Xu1996}. The fluorescence rate is $1/\tau$, where $\tau$ is the fluorescence lifetime. This two-level model is an empirical abstraction of the dynamics of the singlet and triplet states and the intersystem crossing \cite{Gatzogiannis2011,Yonemaru2014}. A rate equation describing this model can be written as
\begin{equation}
	\frac{dN_2(t)}{dt}=g_p\sigma_N\phi^N(t)\left(N_0-N_2(t)\right)-\frac{N_2(t)}{\tau}.
	\label{eq:diff_eq_N2}
\end{equation}
We can convert $\phi(t)$ and $N_2(t)$ to experimentally measurable excitation irradiance, $I(t)$, and fluorescence intensity, $F(t)$, by $I(t)=\phi(t)h c/\lambda$ and $F(t)=N_2(t)\psi_F t_{ob}/\tau$, respectively, where $h$ is Planck's constant, $c$ is the velocity of light, $\lambda$ is the excitation wavelength, $\psi_F$ is the fluorescence detection efficiency, and $t_{ob}$ is the observation time. Denoting $\gamma=\lambda/(h c)$ and $K=\psi_F t_{ob}/\tau$, from Eq. (\ref{eq:diff_eq_N2}), we have
\begin{equation}
	\frac{dF(t)}{dt}=KN_0g_p\sigma_N\gamma^NI^N(t)-\left(g_p\sigma_N\gamma^NI^N(t)+\frac{1}{\tau}\right)F(t).
	\label{eq:diff_eq_F}
\end{equation}

The SOS technique only requires the linear combination of conventional fluorescence intensity images. Therefore it is valid to solely consider steady-state solutions of the fluorophore model. Dropping the temporal dependency, the steady-state solution of Eq. \ref{eq:diff_eq_F} is
\begin{equation}
	F=KN_0\frac{\tau g_p\sigma_N\gamma^NI^N}{1+\tau g_p\sigma_N\gamma^NI^N}.
	\label{eq:steady_state_F}
\end{equation}
Although higher excited singlet and triplet states and photobleaching are not considered in this two-level system, the model can precisely describe the fluorophore saturation behavior. Fig. \ref{fig:1_model}(b) plots the simulated fluorescence-excitation relation described by this model for the 1PEF case. In the range of low excitation irradiance, the fluorescence-excitation relation is linear and unsaturated, while in the range of high excitation irradiance, the relation becomes nonlinear and saturated. Note that the saturation phenomenon can happen without a high excitation intensity. As shown in the inset of Fig. \ref{fig:1_model}(b), even for a moderate excitation irradiance, the simulated curve can deviate from the unsaturated linear curve. We call this phenomenon ``weak saturation'' and it is the basis for the SOS technique. Because of the ``weak saturation'' phenomenon, SOS microscopy is able to produce super-resolution images at a relatively low excitation intensity. Under the ``weak saturation'' condition, denoting $a=\tau g_p \sigma_N \gamma^N$, Eq. \ref{eq:steady_state_F} can be Taylor expanded to
\begin{equation}
	F=KN_0\sum_{n=1}^{\infty}(-1)^{n+1} \tau^ng_p^n\sigma_N^n\gamma^{nN}I^{nN}=KN_0\left(aI^N-a^2I^{2N}+a^3I^{3N}-\cdots\right).
	\label{eq:Taylor_F_expand}
\end{equation}

An $M$-step SOS microscopy needs $M$ fluorescence images to be collected.
Here we consider one-dimensional spatial dependence of excitation and fluorescence. The extension to a multi-dimensional case is trivial. A Gaussian excitation profile is assumed in the focus, $I(x)=I_0 \exp(-2 x^2/\omega_0^2)$ with the focal irradiance $I_0=I(x=0)$ and $1/e^2$ radius $\omega_0$. In the case of $N$-photon excitation, this profile is effectively powered to the $N$-th.
A subscript $i$ is added to the excitation and fluorescence intensities of the $i$-th step image among the $M$ images.
We can separate the spatial dependency by denoting $g(x)=\exp(-2 x^2/\omega_0^2)$. Then for the $i$-th step, we have $I_i(x)=I_{0i} g(x)$ and Eq. \ref{eq:Taylor_F_expand} can be written as
\begin{equation}
	F_i(x)=KN_0\left(aI_{0i}^Ng^N(x)-a^2I_{0i}^{2N}g^{2N}(x)+a^3I_{0i}^{3N}g^{3N}(x)-\cdots\right).
	\label{eq:Fi_expand}
\end{equation}
In Eq. \ref{eq:Fi_expand}, high powers of $g^N(x)$, such as $g^{2N}(x)$, $g^{3N}(x)$, etc., represent components with higher spatial frequency, while $g^N(x)$ is the diffraction-limited component. With the definition of $g(x)$, an $M$-th order power component, $g^{MN}(x)$, has a $\sqrt{M}$-fold increase in spatial resolution. Due to the magnitude difference among each components, the spatial resolution of $F_i(x)$ is dominated by the lowest power of $g^N(x)$.

The idea of an $M$-step SOS microscopy is to eliminate the lowest $M-1$ powers of $g^N(x)$ by the linear combination of $M$ steps of conventional images obtained at different excitation intensities.
We assume the excitation intensities of each step follow $I_{01}<I_{02}<\cdots<I_{0M}$ and they stay in the ``weak saturation'' region. 
The resulting SOS image is $F_{M\textnormal{-}SOS}(x)=\sum_{i=1}^{M}c_iF_i(x)$, where the coefficients $c_i$ are chosen such that the lowest power of $g^N(x)$ in $F_{M\textnormal{-}SOS}(x)$ is $g^{MN}(x)$. 
In other words, an $M$-step SOS image will have a $\sqrt{M}$-fold increase in spatial resolution.

The coefficients of the linear combination, $c_1$, $c_2$, $\cdots$, $c_M$, of the images with excitation intensities, $I_{01}$, $I_{02}$, $\cdots$, $I_{0M}$, are calculated by solving $M$ variables from $M-1$ equations, where the $M$ variables are the coefficients and the $M-1$ equations are the constraints that the lowest $M-1$ powers of $g^N(x)$ in the resulting $M$-step SOS image, $F_{M\textnormal{-}SOS}(x)$, are eliminated. Since this is an under-constrained problem, we set $c_1=1$ without loss of generality. Table \ref{tab:SOS_coefficients} presents the coefficients for the linear combination of SOS methods from two-step to four-step. 
\begin{table}[!t]
	\centering
	\caption{Coefficients for the linear combination in SOS microscopy.}
	\begin{tabular}{ c|c|c }
		\hline
		\multicolumn{2}{c|}{Coefficients} & Expressions \\
		\hline
		\multirow{2}{*}{Two-step} & $c_1$ & $1$ \\
		& $c_2$ & $- \frac{I_{01}^N} {I_{02}^N}$ \\
		\hline
		\multirow{3}{*}{Three-step} & $c_1$ & $1$ \\
		& $c_2$ & $-\frac{I_{01}^{N} (I_{01}^{N} - I_{03}^N)}{I_{02}^N  (I_{02}^N - I_{03}^N)}$ \\
		& $c_3$ & $\frac{I_{01}^{N} (I_{01}^{N} - I_{02}^N)}{I_{03}^N  (I_{02}^N - I_{03}^N)}$ \\ 
		\hline
		\multirow{4}{*}{Four-step} & $c_1$ & $1$ \\
		& $c_2$ & $-\frac{I_{01}^{N} (I_{01}^{N} - I_{03}^N) (I_{01}^{N} - I_{04}^N)}{I_{02}^N  (I_{02}^N - I_{03}^N)  (I_{02}^N - I_{04}^N)}$ \\
		& $c_3$ & $\frac{I_{01}^{N} (I_{01}^{N} - I_{02}^N) (I_{01}^{N} - I_{04}^N)}{I_{03}^N  (I_{02}^N - I_{03}^N)  (I_{03}^N - I_{04}^N)}$ \\ 
		& $c_4$ & $-\frac{I_{01}^{N} (I_{01}^{N} - I_{02}^N) (I_{01}^{N} - I_{03}^N)}{I_{04}^N  (I_{02}^N - I_{04}^N)  (I_{03}^N - I_{04}^N)}$ \\ 
		\hline
	\end{tabular}
	\label{tab:SOS_coefficients}
\end{table}
For instance, in two-step SOS, we need two images, $F_1(x)$ and $F_2(x)$, obtained at excitation intensities, $I_{01}$ and $I_{02}$, respectively. With the power series in Eq. \ref{eq:Fi_expand} and the two-step coefficients, $c_1$ and $c_2$, in Tab. \ref{tab:SOS_coefficients}, the resulting two-step SOS image is $F_{2\textnormal{-}SOS}(x)=KN_0[-a^2I_{01}^{N}(I_{01}^{N}-I_{02}^{N})g^{2N}(x)+a^3I_{01}^{N}(I_{01}^{2N}-I_{02}^{2N})g^{3N}(x)-\cdots]$, 
which is dominated by the component $g^{2N}(x)$.
Compared to the diffraction-limited component $g^{N}(x)$, $F_{2\textnormal{-}SOS}(x)$ has a $\sqrt{2}$-fold increase in spatial resolution.
In general, for an $M$-step SOS, a super-resolved image with a $\sqrt{M}$-fold resolution improvement compared to the diffraction limit can be achieved by the linear combination with the coefficients in Tab. \ref{tab:SOS_coefficients}. Although Tab. \ref{tab:SOS_coefficients} only provides SOS coefficients from two-step to four-step due to limited space for this article, the coefficients for any number of steps can be calculated using the methods described above.
Alternatively, by inspecting the mathematical expressions in Tab. \ref{tab:SOS_coefficients}, a rule about the formation of SOS coefficients can be observed. Specifically, for two adjacent steps $A$ and $B$, where $B=A+1$, if $i<B$, the coefficient $c_i$ in $B$-step SOS is the product of $c_i$ in $A$-step SOS and a factor $(I_{01}^N-I_{0B}^N)/(I_{0i}^N-I_{0B}^N)$; if $i=B$, then $c_i$ in $B$-step SOS is
\begin{equation}
	c_B=(-1)^{B-1}\frac{I_{01}^N}{I_{0B}^N}\prod_{j=2}^{B-1}\frac{I_{01}^N-I_{0j}^N}{I_{0j}^N-I_{0B}^N}.
	\label{eq:last_coefficient_rule}
\end{equation}

\begin{figure*}[!t]
	\centering
	\includegraphics[width=0.9\linewidth]{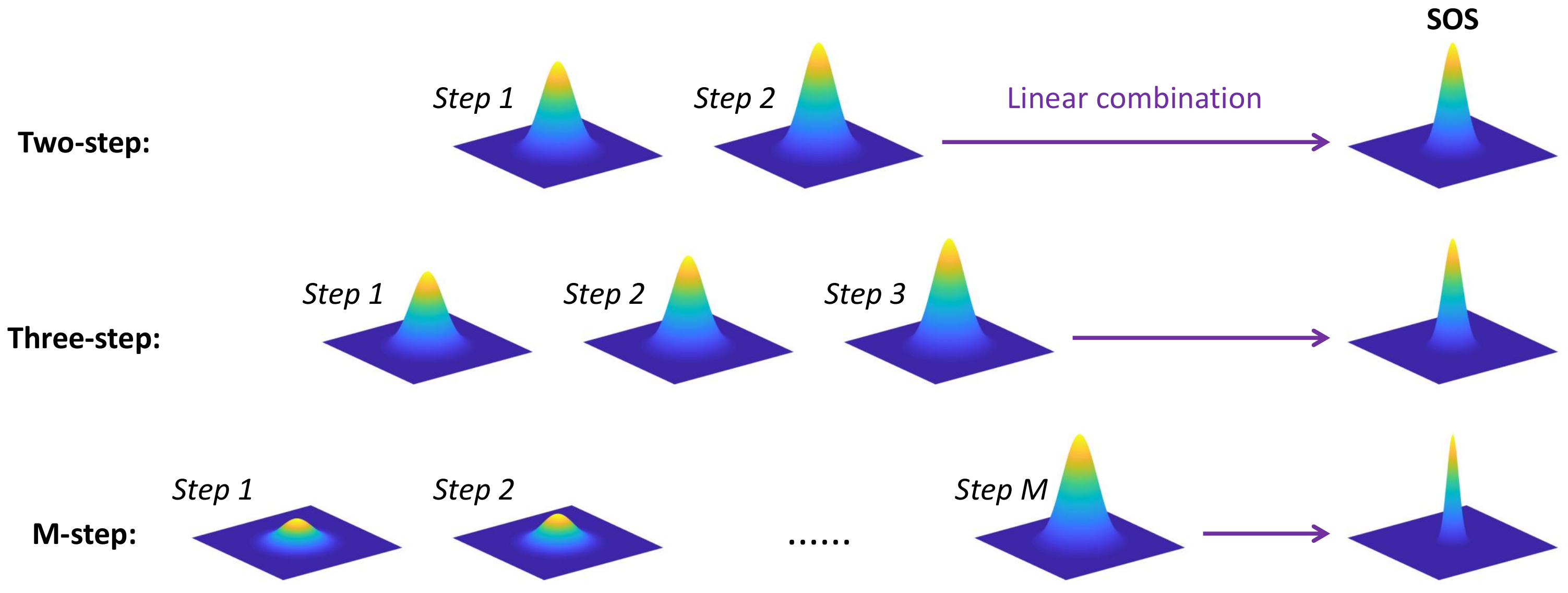}
	\caption{Illustration of the principle of SOS microscopy from two-step to $M$-step.}
	\label{fig:2_illustration}
\end{figure*}
Figure \ref{fig:2_illustration} graphically illustrates how super-resolution images are obtained in SOS microscopy. Generally, in an $M$-step microscopy, $M$ steps of diffraction-limited raw images [here we only plot point spread functions (PSFs) for illustration] obtained at $M$ different excitation intensities, $I_{01}$, $I_{02}$, $\cdots$, $I_{0M}$, are linearly combined to produce a super-resolution SOS image with a $\sqrt{M}$-fold increase in spatial resolution (a PSF $\sqrt{M}$-fold narrower than a diffraction-limited PSF). The linear combination coefficients, $c_1$, $c_2$, $\cdots$, $c_M$, are functions of the intensities and are presented in Tab. \ref{tab:SOS_coefficients}.
For instance, in two-step SOS, we need two images, $F_1(x)$ and $F_2(x)$, obtained at excitation intensities, $I_{01}$ and $I_{02}$, respectively. With the two-step coefficients, $c_1$ and $c_2$, in Tab. \ref{tab:SOS_coefficients}, the resulting two-step SOS image $F_{2\textnormal{-}SOS}(x)=c_1F_1(x)+c_2F_2(x)$ is dominated by the component $g^{2N}(x)$. Compared to the diffraction-limited component $g^{N}(x)$, $F_{2\textnormal{-}SOS}(x)$ has a $\sqrt{2}$-fold increase in spatial resolution. 
Similarly, three-step SOS needs three raw images and the resulting SOS image, $F_{3\textnormal{-}SOS}(x)$, is dominated by the components $g^{3N}(x)$, which has a $\sqrt{3}$-fold increase in spatial resolution. 
The same principle applies to $M$-step SOS, which is able to generate a super-resolution image with a $\sqrt{M}$-fold increase in spatial resolution.

\section{Numerical Results}
\label{sec:numerical}

\begin{figure}[!t]
	\centering
	\includegraphics[width=0.8\linewidth]{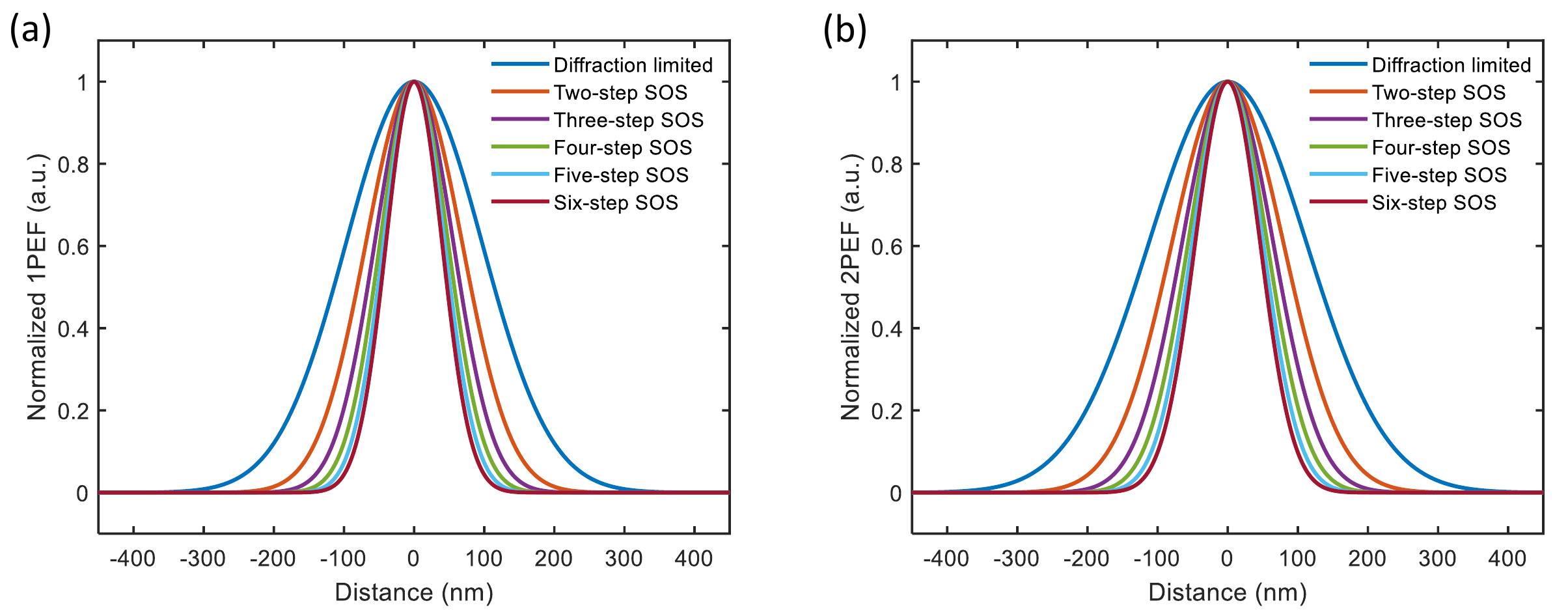}
	\caption{PSFs of SOS microscopy from two-step to six-step for (a) 1PEF and (b) 2PEF, where diffraction-limited PSFs are also plotted for comparison.}
	\label{fig:3_PSFs}
\end{figure}

The super-resolution capability of SOS microscopy is first investigated using numerical simulations. The simulation is based on the two-level model described above and the parameters are given in Section \ref{sec:materials_methods}. 
Here several diffraction-limited PSFs obtained at different excitation intensities are used to generate SOS images from two-step to six-step. The resulting PSFs of the SOS microscopy for both 1PEF and 2PEF are shown in Fig. \ref{fig:3_PSFs}. 
For the 1PEF case, the conventional diffraction-limited PSF has a full-width at half-maximum (FWHM) of 228.9 nm, while the two-step to six-step SOS's FWHMs are 162.1 nm, 132.5 nm, 114.9 nm, 103.1 nm, and 94.5 nm, corresponding to a 1.41, 1.73, 2.00, 2.22, and 2.42 folds increase in resolution, respectively.
For the 2PEF case, the conventional diffraction-limited FWHM is 265.3 nm, while the FWHMs of the two-step to six-step SOS are 187.7 nm, 153.5 nm, 133.3 nm, 119.5 nm, and 109.5 nm, which, identically, correspond to a 1.41, 1.73, 2.00, 2.22, and 2.42 folds increase in resolution, respectively.
It is clearly shown that the resolution improvements for both 1PEF and 2PEF are equivalent, and for an $M$-step SOS, whether it is 1PEF or 2PEF, the resolution improvement is exactly $\sqrt{M}$-fold.
\begin{figure}[!t]
	\centering
	\includegraphics[width=0.9\linewidth]{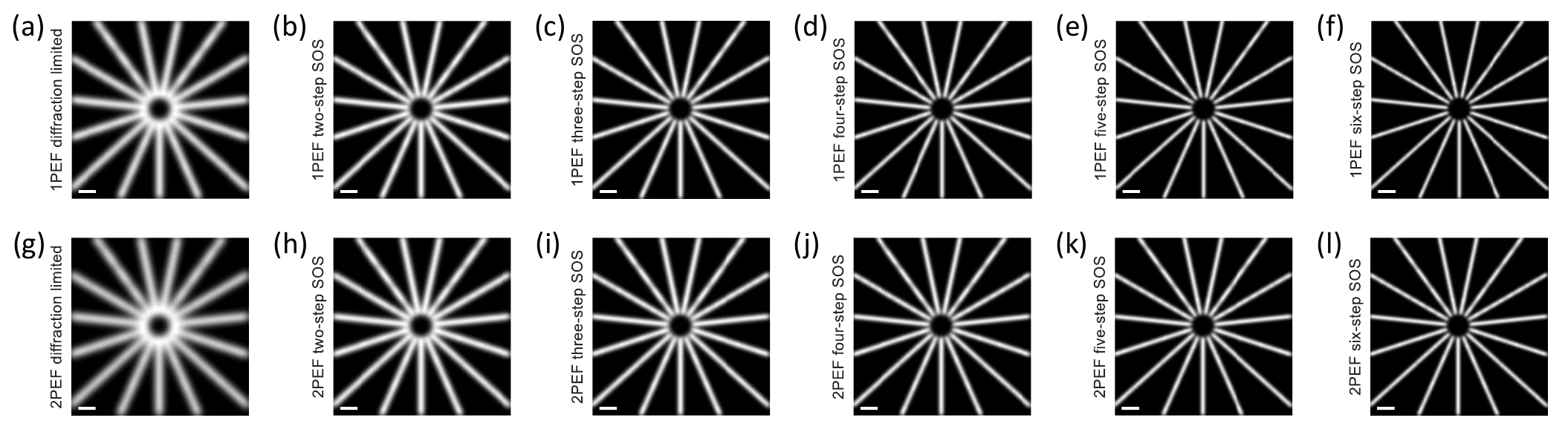}
	\caption{Simulated SOS images of a two-dimensional artificial object. Scale bar: 500 nm.}
	\label{fig:4_targets}
\end{figure}

Although the principle of SOS microscopy is derived with one-dimensional objects. The extension to two- or multi-dimensional cases is trivial. Here we apply the SOS methods to a two-dimensional simulation target, as shown in Fig. \ref{fig:4_targets}. The target is a star-like object consisting of several lines, where the closet distance of the lines is 140 nm. The PSFs in Fig. \ref{fig:3_PSFs} are used as kernels to simulate the diffraction-limited and SOS images for both 1PEF and 2PEF cases. The diffraction-limited images for both cases are blurred and the center ends of the lines cannot be differentiated. With the help of the SOS microscopy, super-resolved images of the object can be obtained. In the simplest two-step SOS, an obvious resolution improvement can be seen for both 1PEF and 2PEF. As the number of SOS steps increases, the resolution improvement is more significant. The performance of the images in Fig. \ref{fig:4_targets} matches well with the FWHM results in Fig. \ref{fig:3_PSFs}, as the center ends of the lines in the object are successfully differentiated in three-step to six-step 1PEF SOS [Figs. \ref{fig:4_targets}(c-f)] and four-step to six-step 2PEF SOS [Figs. \ref{fig:4_targets}(j-l)], whose FWHMs are smaller than the closet distance of the lines (140 nm) in the object.

\section{Experimental Results}
\label{sec:experimental}

\begin{figure}[!t]
	\centering
	\includegraphics[width=0.9\linewidth]{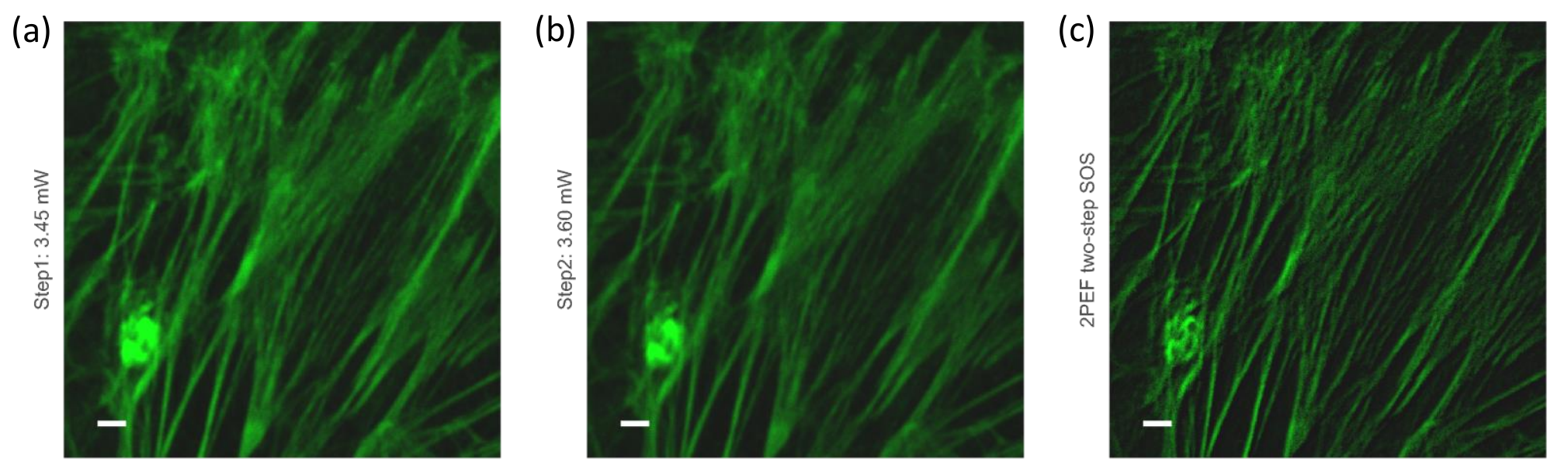}
	\caption{Representative example of 2PEF two-step SOS with Alexa Fluor 488 phalloidin labeled F-actin in fixed bovine pulmonary artery endothelial (BPAE) cells. (a) The raw image of step1 with an MPM laser power of 3.45 mW. (b) The raw image of step2 with an MPM laser power of 3.60 mW. (c) The processed 2PEF two-step SOS image using the raw images in (a) and (b). Scale bar: 2 $\upmu$m.}
	\label{fig:5_exp_example}
\end{figure}
We now present the experimental validation of the proposed SOS methods. Although SOS microscopy can provide a $\sqrt{M}$-fold increase in spatial resolution with the linear combination of $M$ steps of conventional images, and theoretically this resolution improvement can be infinite, practically, however, an experimental SOS image generated from more than two steps of raw images is generally unacceptable due to its extremely poor SNR.
For example, for two steps of 1PEF images obtained with irradiances $I$ and $1.05I$, the two-step coefficients, based on Tab. \ref{tab:SOS_coefficients}, are $c_1=1$ and $c_2=-0.952$; for three-step 1PEF images with irradiances $I$, $1.05I$, and $1.1I$, the coefficients are $c_1=1$, $c_2=-1.905$, and $c_3=0.909$, respectively. Since the excitation irradiances for each image are approximate, if we assume the fluorescence intensities of raw images are the same and equal to $F$, we can estimate the intensity of the resulting SOS image after the linear combination, $\sum_{i=1}^{M}c_iF$. Hence, the intensities of two-step and three-step SOS images will be $(c_1+c_2)F=0.048F$ and $(c_1+c_2+c_3)F=0.004F$, respectively, which are one order and two orders of magnitude lower compared with the raw intensity $F$.
Considering the shot noise and the propagation of errors in SOS algorithms \cite{Zhang2016}, the SNR of a two-step SOS image will be an order of magnitude lower than a conventional image, while for SOS images generated from more than two steps, the SNR will be at least two orders of magnitude lower.
Therefore in the experiments we only present confocal (1PEF) and multiphoton (2PEF) two-step SOS images for their satisfactory SNR performance and easy implementation.
Figure \ref{fig:5_exp_example} shows a representative example of how two raw images of Alexa Fluor 488 phalloidin labeled F-actin in fixed cells are obtained and processed to generate the super-resolved two-step SOS image. 
The raw images in Figs. \ref{fig:5_exp_example}(a) and \ref{fig:5_exp_example}(b) were obtained with Nikon's commercial software.
These two images were then imported into Matlab and converted to matrices $\mathbf{M_1}$ and $\mathbf{M_2}$. 
Based on the knowledge of the excitation powers of these two images, the corresponding two-step SOS image was computed using the coefficients shown in Tab. \ref{tab:SOS_coefficients}. 
In the 2PEF two-step SOS case of Fig. \ref{fig:5_exp_example}, we have $N=2$, $I_{01}=3.45$ mW, and $I_{02}=3.60$ mW, so the SOS coefficients are $c_1=1$ and $c_2=-I_{01}^N/I_{02}^N=-0.92$ and the resulting SOS image is $\mathbf{M_{2\textnormal{-}SOS}}=c_1\mathbf{M_1}+c_2\mathbf{M_2}=\mathbf{M_1}-0.92\mathbf{M_2}$. The resolution improvement of the two-step SOS image is appreciable.
The materials and methods used in the experiments in this Section are described in Section \ref{sec:materials_methods}.

\begin{figure}[!t]
	\centering
	\includegraphics[width=0.8\linewidth]{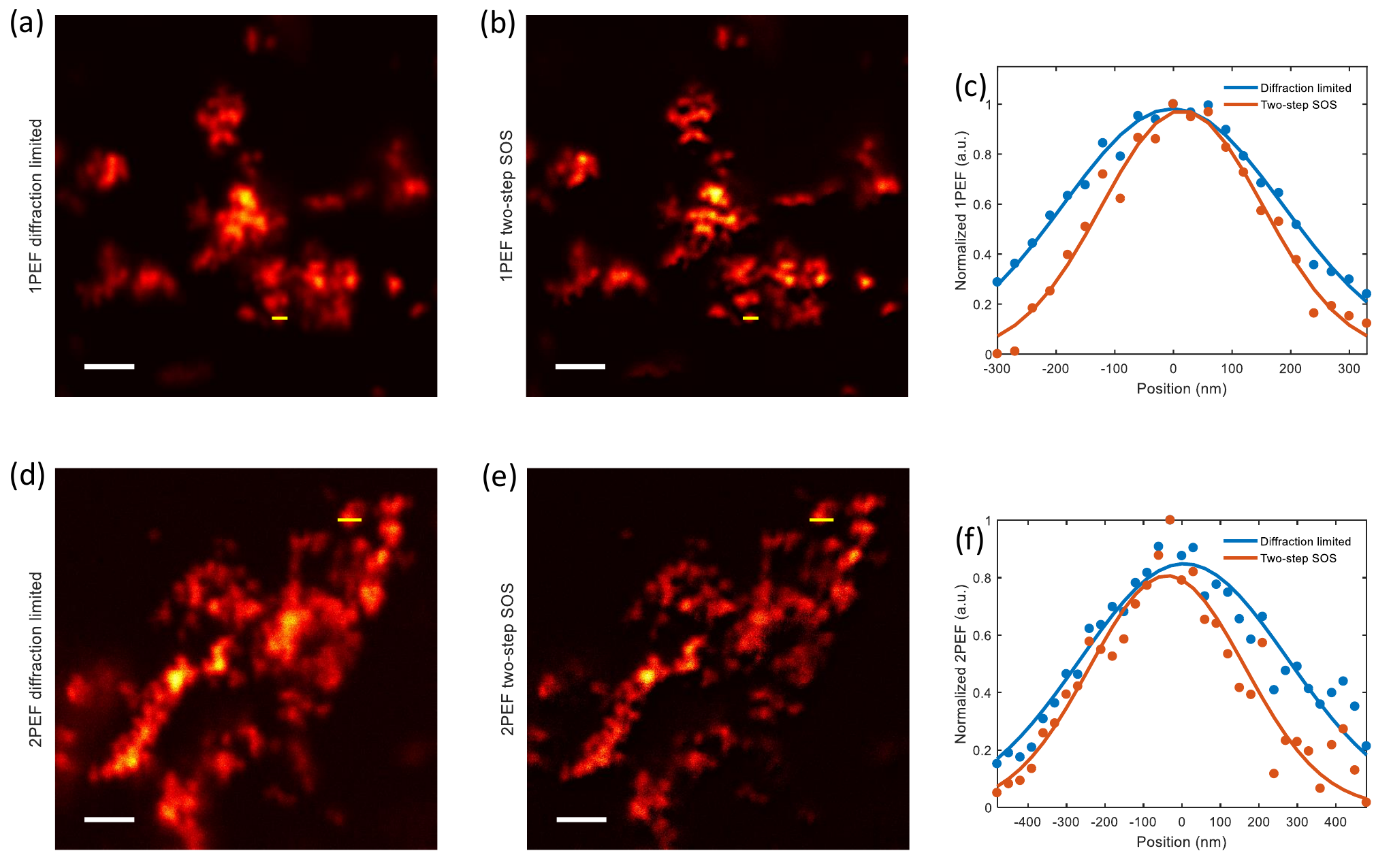}
	\caption{Experimental SOS images of RITC-$\mathrm{SiO_2}$-SiEDTA NPs in an agarose gel. The confocal (1PEF) (a) diffraction-limited and (b) two-step SOS images of the sample. (c) Intensity profiles (dots) with Gaussian fits (curves) along the yellow lines in (a) and (b). The multiphoton (2PEF) (d) diffraction-limited and (e) two-step SOS images of the sample. (f) Intensity profiles (dots) with Gaussian fits (curves) along the yellow lines in (d) and (e). Scale bar: 2 $\upmu$m.}
	\label{fig:6_NPgel}
\end{figure}
Next, we imaged a phantom sample consisting of subdiffractive fluorescent nanoparticles (NPs) embedded in a hydrogel matrix (0.25\%, \textasciitilde55 nm RITC-$\mathrm{SiO_2}$-SiEDTA NPs dispersed in 1\% agarose gel). The conventional and two-step SOS images of the sample obtained with 1PEF and 2PEF are presented in Fig. \ref{fig:6_NPgel}. Compared to conventional, diffraction-limited 1PEF and 2PEF images in Figs. \ref{fig:6_NPgel}(a) and \ref{fig:6_NPgel}(d), the 1PEF and 2PEF two-step SOS images in Figs. \ref{fig:6_NPgel}(b) and \ref{fig:6_NPgel}(e) are super-resolved where NPs at different sites are better differentiated. The intensity profiles of a cluster of NPs obtained with 1PEF and 2PEF are fitted to Gaussian curves and plotted in Figs. \ref{fig:6_NPgel}(c) and \ref{fig:6_NPgel}(f), respectively. For the 1PEF case, the diffraction-limited Gaussian curve has a FWHM of 442.2 nm, while the two-step SOS's FWHM is 325.1 nm, corresponding to a 1.36-fold increase in resolution. For the 2PEF case, the diffraction-limited FWHM is 636.9 nm, while the FWHM of the two-step SOS is 474.9 nm, which corresponds to a 1.34-fold increase in resolution. The resolution improvements in both 1PEF and 2PEF two-step SOS microscopy are close to the theoretical value, $\sqrt{2}=1.41$. The discrepancy, which is more severe for the 2PEF case, is due to the limited SNR of the experimentally obtained images.

\begin{figure}[!t]
	\centering
	\includegraphics[width=0.8\linewidth]{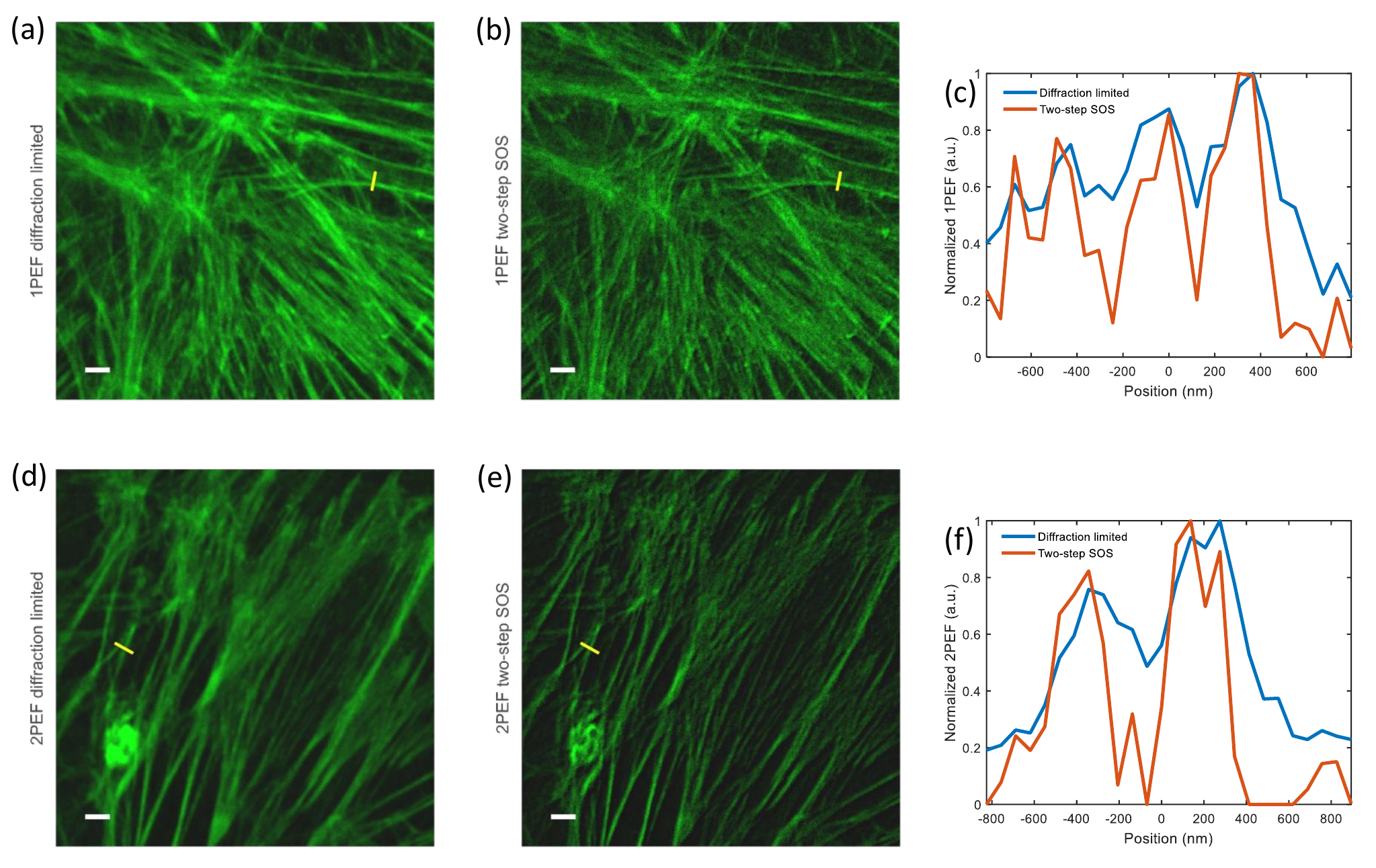}
	\caption{Experimental SOS images of the Alexa Fluor 488 phalloidin labeled F-actin in the biological test slide (fixed BPAE cells, FluoCells prepared slide \#1, F36924). The confocal (1PEF) (a) diffraction-limited and (b) two-step SOS images. (c) Intensity profiles along the yellow lines in (a) and (b). The multiphoton (2PEF) (d) diffraction-limited and (e) two-step SOS images. (f) Intensity profiles along the yellow lines in (d) and (e). Scale bar: 2 $\upmu$m.}
	\label{fig:7_test}
\end{figure}
We further studied the resolution improvement enabled by two-step SOS microscopy using a standard biological test slide [F-actin labeled with Alexa Fluor 488 phalloidin in fixed BPAE cells, FluoCells prepared slide \#1, F36924] \cite{Zucker2006}. 
Figure \ref{fig:7_test} shows the conventional and SOS images of the F-actin filaments with both 1PEF and 2PEF. By comparing the conventional [Figs. \ref{fig:7_test}(a) and \ref{fig:7_test}(d)] with the two-step SOS [Figs. \ref{fig:7_test}(b) and \ref{fig:7_test}(e)] images, one can see the significant sharpening of the actin structures enabled by the super-resolution SOS method in both 1PEF and 2PEF imaging. The intensity profiles of a few closely located filaments shown in Figs. \ref{fig:7_test}(c) and \ref{fig:7_test}(f) demonstrate the finer apparent width and better resolution of the filaments in two-step SOS images. In conventional 1PEF and 2PEF images, the differentiation of these filaments is not possible.

\begin{figure}[!t]
	\centering
	\includegraphics[width=0.8\linewidth]{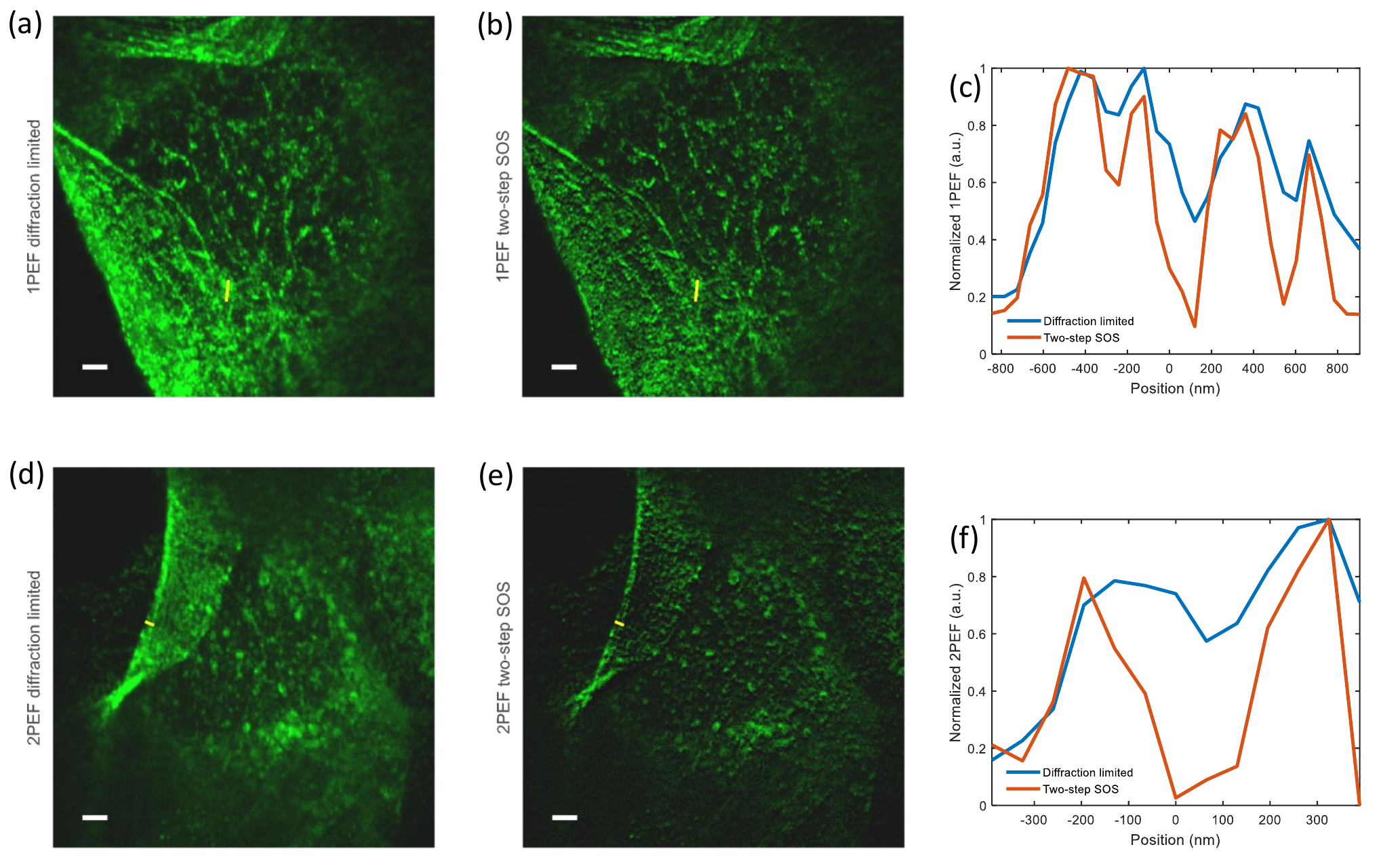}
	\caption{Experimental SOS images of the Alexa Fluor 488 phalloidin labeled F-actin in the fixed cells (ECFCs) sample. The confocal (1PEF) (a) diffraction-limited and (b) two-step SOS images. (c) Intensity profiles along the yellow lines in (a) and (b). The multiphoton (2PEF) (d) diffraction-limited and (e) two-step SOS images. (f) Intensity profiles along the yellow lines in (d) and (e). Scale bar: 2 $\upmu$m.}
	\label{fig:8_cells}
\end{figure}
Having verified the resolution improvement with a standard sample, we then validated the SOS methods using a biological sample [F-actin labeled with Alexa Fluor 488 phalloidin in fixed endothelial colony forming cells (ECFCs)] prepared in our labs. The conventional and two-step imaging results for 1PEF and 2PEF are shown in Fig. \ref{fig:8_cells}. Since the fluorophore concentration used to stain this sample is lower than the one used in the standard test slide (Fig. \ref{fig:7_test}), the fluorophores, though successfully bind to F-actin, look like subdiffractive beads when imaged. The improvement in resolution and imaging quality of two-step SOS microscopy [Figs. \ref{fig:8_cells}(b) and \ref{fig:8_cells}(e)] can be easily seen compared with the conventional diffraction-limited 1PEF and 2PEF images [Figs. \ref{fig:8_cells}(a) and \ref{fig:8_cells}(d)]. The intensity profiles in Figs. \ref{fig:8_cells}(c) and \ref{fig:8_cells}(f) show that, in both 1PEF and 2PEF imaging, the neighboring conventionally unresolvable fluorophores can be clearly resolved using two-step SOS microscopy.

\begin{figure}[!t]
	\centering
	\includegraphics[width=0.9\linewidth]{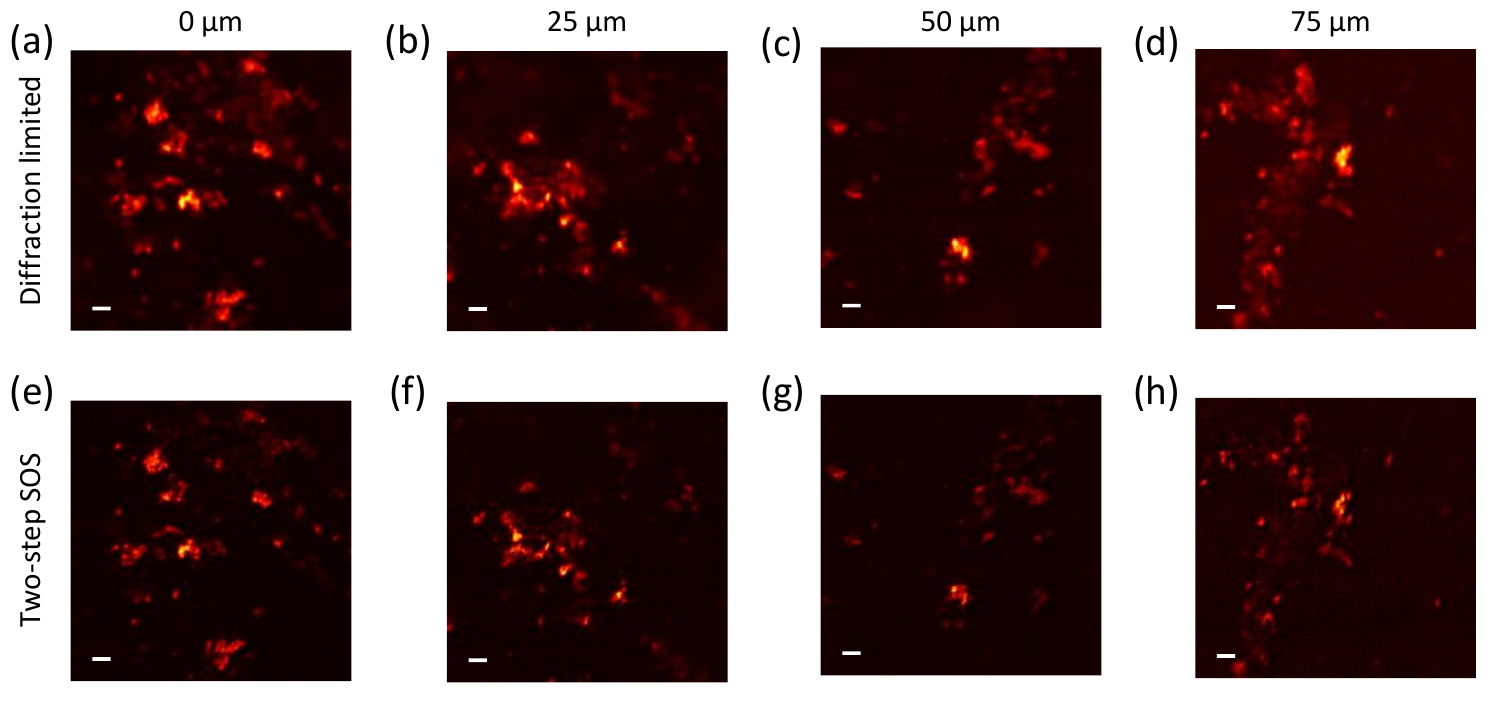}
	\caption{Experimental multiphoton (2PEF) SOS images of the RITC-$\mathrm{SiO_2}$-SiEDTA NPs in a scattering phantom mimicking a brain tissue at various depths. ``0 $\upmu$m'' corresponds to the coverslip surface. The 2PEF (a-d) diffraction-limited and (e-h) two-step SOS images of the sample at depths of 0 $\upmu$m, 25 $\upmu$m, 50 $\upmu$m, and 75 $\upmu$m from the coverslip. Scale bar: 2 $\upmu$m.}
	\label{fig:9_depth}
\end{figure}
Since two-step SOS microscopy was able to generate super-resolution multiphoton (2PEF) images, we then investigated the advantages of 2PEF, such as deep penetration and confined photodamage, in addition to super-resolution, in 2PEF two-step SOS microscopy.
We used the same subdiffractive fluorescent NPs used in Fig. \ref{fig:6_NPgel} (0.25\%, \textasciitilde55 nm RITC-$\mathrm{SiO_2}$-SiEDTA NPs) but dispersed in a scattering phantom also comprising polystyrene beads within the agarose gel to mimic the optical scattering inside a brain tissue \cite{Yoshida2015}. The experimental 2PEF conventional and two-step SOS images at various depths from the coverslip surface are shown in Fig. \ref{fig:9_depth}. By virtue of the deep penetration of 2PEF, for a diffraction-limited 2PEF image obtained at each depth (from 0 $\upmu$m to 75 $\upmu$m) in a scattering phantom, we are able to extract a corresponding super-resolved two-step SOS image where conventionally unresolvable NPs can be clearly resolved.
In principle, as long as a raw fluorescence image with an acceptable SNR can be obtained, a corresponding super-resolution two-step SOS image can be generated, regardless of the experimental conditions such as imaging depth.

\section{Materials and Experimental Methods}
\label{sec:materials_methods}
\subsection{Materials}
\label{subsec:materials}
Alexa Fluor-488 conjugated phalloidin (Invitrogen), Ammonium hydroxide ($\mathrm{NH_4OH}$, 20\%, VWR), (3-aminopropyl) triethoxysilane (APTES, $\mathrm{C_9H_{23}NO_3Si}$, >98\%, Sigma, St. Louis, MO), Aqua-Mount (Lerner laboratories), endothelial growth medium (EBM-2, Lonza), ethanol (200 proof, VWR), fetal bovine serum (FBS, GE life sciences), 10\% w/v 1.1 micron Latex beads, polystyrene (PS beads, Sigma), 1000U/mL penicillin-streptomyocin (Corning), rhodamine B isothiocyanate (RITC, $\mathrm{C_{29}H_{30}ClN_3O_3S}$, Sigma), tetraethylorthosilicate (TEOS, $\mathrm{C_8H_{20}O_4Si}$, >99\%, Sigma), triton-x-100 ($\mathrm{C_{14}H_{22}O}$, Sigma-Aldrich) and N-(trimethoxysillylpropyl)ethylenediaminetriacete, sodium salt, 30\% in water (SiEDTA, $\mathrm{C_{14}H_{25}N_2Na_3O_9Si}$,  GELEST) were all used as-received.

\subsection{Dispersion of 55 nm Fluorescent Silica Nanoparticles in Agarose Gel and Brain Optical Phantoms}
\label{subsec:55nmNP_gel}
A stock solution of silanized RITC was obtained by reacting RITC with APTES in a 1:1 mole ratio, overnight (RITC-APTES). Fluorescent silica nanoparticles (NPs) were synthesized by magnetically stirring 10 ml of ethanol and 2 ml of water in a 20 mL glass vial.  0.3 mL of TEOS and 0.075 mL of RITC-APTES were added sequentially and the solution was left stirring. 0.6 mL of 20\% w/v $\mathrm{NH_4OH}$ was then added rapidly to the ethanolic solution of TEOS/RITC-APTES. The mixture was stirred for 12 h at room temperature, in the dark. The fluorescent silica particles (RITC-$\mathrm{SiO_2}$ NPs) were recovered by centrifugation at 12000 relative centrifugal force (rcf) for 45 min, discarding the supernatant and resuspending the pellet in ethanol via ultrasonication. The as-prepared RITC-$\mathrm{SiO_2}$ NPs were amphiphilic and could be dispersed in either ethanol or water.
10 mL of a 1\% w/v solution of RITC-$\mathrm{SiO_2}$ NPs in 80/20 ethanol/water solution was vigorously stirred with 0.1 mL of 30\% w/v SiEDTA solution for 24 h in the dark, modifying previously established methods \cite{Nallathamby2016,Nallathamby2015,Wang2013}. The silanized RITC-$\mathrm{SiO_2}$-SiEDTA NPs were rinsed thrice in 50/50 ethanol/water by centrifugation at 12000 rcf for 45 min and discarding the supernatant. After the final rinse, the RITC-$\mathrm{SiO_2}$-SiEDTA NPs were resuspended in de-ionized water via ultrasonication.
RITC-$\mathrm{SiO_2}$-SiEDTA NPs were prepared at 0.5\%, 0.25\% and 0.125\% w/v in 1\% agarose by mixing 500 $\upmu$L, 250 $\upmu$L and 125 $\upmu$L of 1\% w/v RITC-$\mathrm{SiO_2}$-SiEDTA NPs with 500 $\upmu$L, 750 $\upmu$L and 875 $\upmu$L of a warm 2\% agarose solution, respectively. 100 $\upmu$L of the RITC-$\mathrm{SiO_2}$-SiEDTA NPs in 1\% agarose were deposited on glass slides and the agarose was allowed to spread and set at room temperature.
Additionally, 37 $\upmu$L of 10\% w/v PS beads were added to each mixture. 100 $\upmu$L of the mixture of PS beads and RITC-$\mathrm{SiO_2}$-SiEDTA NPs in 1\% agarose were then spread on glass slides and the agarose was allowed to set at room temperature. The mixture of fluorescent silica particles, agarose and polystyrene beads mimicked the optical scattering inside a brain tissue \cite{Yoshida2015}.

\subsection{Simulation Parameters}
\label{subsec:simulation_params}

The simulation parameters in this research are extracted from previous works \cite{Eggeling2005,Zhang2017}. Unless otherwise noted, all parameters are set as follows:
$\mathrm{NA}=0.8$, $\tau=3$ ns, $t_{ob}=35.2$ $\upmu$s, $\psi_F=0.02$, $N_0=1$, $\omega_0=\lambda/(\pi \mathrm{NA})$, $h=6.626\times 10^{-34}$ $\mathrm{J\cdot s}$, $c=3\times 10^{8}$ $\mathrm{m/s}$.
For the simulations with 1PEF, we use
$N=1$, $\lambda=488$ nm, $g_p=1$, $\sigma_1=1.35\times 10^{-20}$ $\mathrm{m}^2$.
For the simulations with 2PEF, we use
$N=2$, $\lambda=800$ nm, $g_p=38690$, $\sigma_2=2\times 10^{-56}$ $\mathrm{m}^4\mathrm{s}$.
In Figs. \ref{fig:3_PSFs} and \ref{fig:4_targets}, the excitation irradiances of each step for 1PEF are $I_{01}=0.22\times 10^{8}$ $\mathrm{W/m^2}$, $I_{02}=0.79\times 10^{8}$ $\mathrm{W/m^2}$, $I_{03}=1.08\times 10^{8}$ $\mathrm{W/m^2}$, $I_{04}=1.70\times 10^{8}$ $\mathrm{W/m^2}$, $I_{05}=4.34\times 10^{8}$ $\mathrm{W/m^2}$, $I_{06}=6.89\times 10^{8}$ $\mathrm{W/m^2}$, and the excitation irradiances for 2PEF are $I_{01}=0.72\times 10^{10}$ $\mathrm{W/m^2}$, $I_{02}=0.94\times 10^{10}$ $\mathrm{W/m^2}$, $I_{03}=1.80\times 10^{10}$ $\mathrm{W/m^2}$, $I_{04}=2.76\times 10^{10}$ $\mathrm{W/m^2}$, $I_{05}=3.04\times 10^{10}$ $\mathrm{W/m^2}$, $I_{06}=4.15\times 10^{10}$ $\mathrm{W/m^2}$, where the $M$-step SOS uses irradiances from $I_{01}$ to $I_{0M}$.

\subsection{Optical System}
\label{subsec:optical}
Both confocal and MPM imaging were performed on a Nikon A1R-MP confocal system. All samples were imaged using a 100x, 1.45 NA, oil-immersion objective (Nikon Plan Apo) except for the NPs in brain phantom sample, which was imaged using a 40x, 1.15 NA, water-immersion objective (Nikon Apo LWD). The excitation in the confocal imaging was generated by a LU4/LU4A laser unit.
The samples with 1PEF were excited at a wavelength of 488 nm (biological test slide and fixed cells sample) and 561 nm (NPs in gel sample). The pinhole size in all confocal imaging was set to 1.2 AU.
The excitation in the MPM was generated by a Spectra-Physics Mai Tai DeepSee femtosecond laser. All samples with 2PEF were excited at a wavelength of 800 nm.
The laser power levels for both confocal and MPM systems were measured at the sample plane with a digital optical power meter (Thorlabs PM100D) and a S120C detector (Thorlabs S120C).
A careful power calibration for both confocal and MPM lasers was performed to correspond laser power settings (\%) with actual focal power measurements (mW). This calibration was required in order to correctly apply the two-step SOS method because the linear combination coefficients in SOS microscopy are based on excitation powers.

\subsection{Image Acquisition and Processing}
\label{subsec:acquisition}
All images were 512x512 pixels and were line-averaged for 16 times for the best SNR.
As two-step SOS microscopy requires a linear combination of two images, for both confocal and MPM imaging, each sample was imaged twice at two different laser powers to generate two raw images.
For the NPs in gel sample (Fig. \ref{fig:6_NPgel}), the confocal images were obtained with laser powers of 1.0\% (3.7 $\upmu$W) and 1.3\% (4.6 $\upmu$W), the PMT HV gain of 110, and the pixel time of 12.1 $\upmu$s; the MPM imaging was performed with laser powers of 1.0\% (3.45 mW) and 1.2\% (3.6 mW), the PMT HV gain of 140, and the pixel time of 12.1 $\upmu$s.
For the biological test slide (Fig. \ref{fig:7_test}), the confocal images were obtained with laser powers of 1.0\% (2.8 $\upmu$W) and 1.2\% (3.2 $\upmu$W), the PMT HV gain of 125, and the pixel time of 4.8 $\upmu$s; the MPM imaging was performed with laser powers of 1.0\% (3.45 mW) and 1.2\% (3.6 mW), the PMT HV gain of 125, and the pixel time of 12.1 $\upmu$s.
For the fixed cells sample (Fig. \ref{fig:8_cells}), the confocal images were obtained with laser powers of 1.0\% (2.8 $\upmu$W) and 1.2\% (3.2 $\upmu$W), the PMT HV gain of 125, and the pixel time of 12.1 $\upmu$s; the MPM imaging was performed with laser powers of 1.0\% (3.45 mW) and 1.2\% (3.6 mW), the PMT HV gain of 140, and the pixel time of 12.1 $\upmu$s.
For the NPs in brain phantom sample (Fig. \ref{fig:9_depth}), the MPM imaging at various depth was performed, with laser powers of 1.0\% (3.45 mW) and 1.2\% (3.6 mW), the PMT HV gain of 130, and the pixel time of 12.1 $\upmu$s.

\section{Discussion}
\label{sec:discussion}

\begin{figure}[!t]
	\centering
	\includegraphics[width=0.8\linewidth]{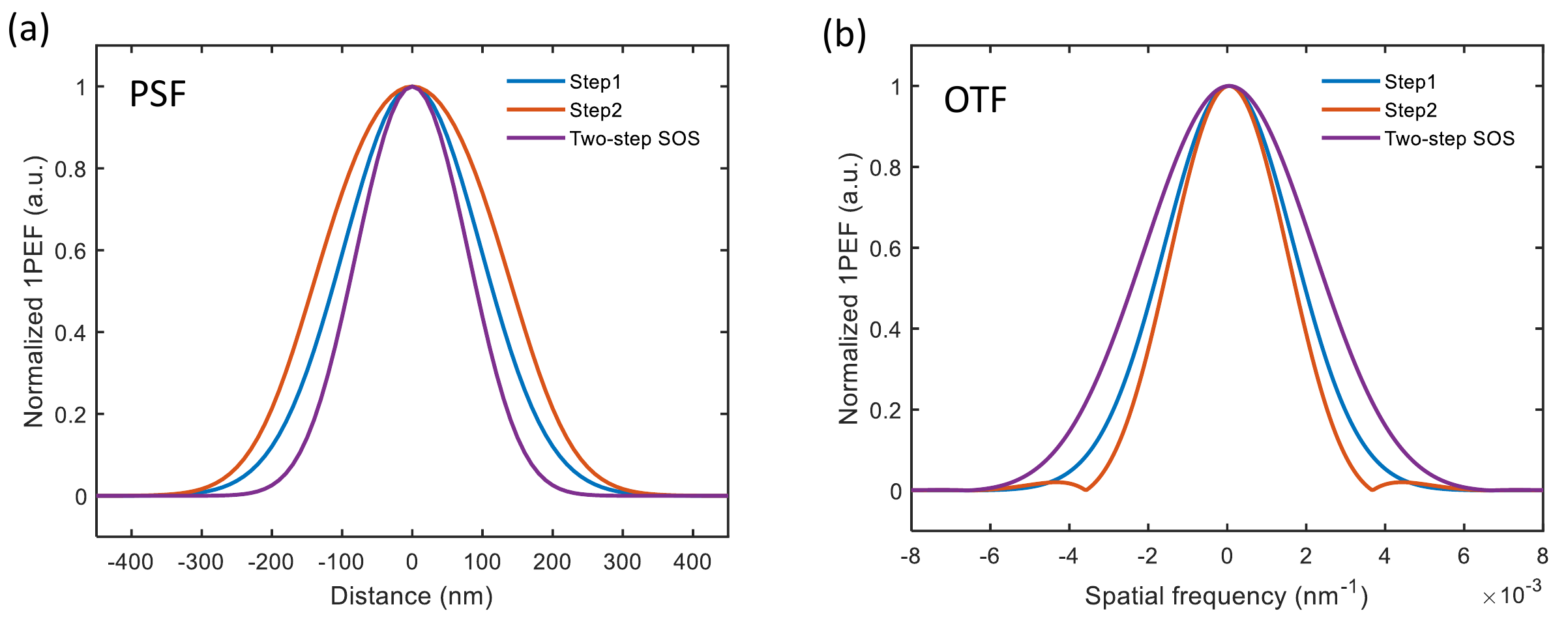}
	\caption{(a) PSFs of two steps of images obtained at different excitation irradiances and their processed two-step SOS image for 1PEF fluorophores. (b) OTFs of corresponding PSFs in (a).}
	\label{fig:S2}
\end{figure}

The resolution improvement in SOS microscopy is observed from the narrowing of its PSF. 
The improvement is fundamentally resulted from the broadening, or the increased FWHM, of its optical transfer function (OTF), which is the Fourier transform of the PSF and the indicator of how much spatial frequency content the image can support.
To demonstrate the broadening of spatial frequency content, we plot PSFs and corresponding OTFs of two steps of images obtained at different excitation intensities and their processed two-step SOS image in Fig. \ref{fig:S2}.
From the first step image to the second, the excitation intensity is increased; therefore the PSF is broadened due to the saturation behavior. Nevertheless, the processed two-step SOS image has a PSF narrower than both of them, as shown in Fig. \ref{fig:S2}(a). Correspondingly, the OTF of the first step image is a Gaussian function with a limited frequency bandwidth, while the OTF of the second step brings about an evident side lobe around the main lobe, as shown in Fig. \ref{fig:S2}(b). In combination with the main lobe, the second step's side lobe contributes to an ``increased'' frequency bandwidth compared to the first step, although the ``increase'' of bandwidth here is not regular as there are zeros in the spatial frequency band. 
With the linear combination algorithm in two-step SOS microscopy, the OTF of the processed two-step SOS image is a single-lobed Gaussian function, which has the largest FWHM and the highest spatial resolution. Note that the zeros in the second step image's OTF is eliminated in the two-step SOS image due to the linear combination such that the ``increase'' of spatial frequency bandwidth here in SOS microscopy is in the regular sense.
This strategy is different than deconvolution because deconvolution cannot increase the spatial frequency bandwidth of an image.

Compared with established super-resolution methods including STED/RESOLFT, PALM, STORM, and SIM, the implementation of SOS microscopy is very easy, fast and requires no new instrumentation. Only two raw images are required to generate a super-resolution SOS image with a $\sqrt{2}$-fold increase in spatial resolution, and theoretically, the resolution improvement can be infinite. 
However, the super-resolution capability of SOS microscopy is limited by a sacrifice in SNR.
In practice, an experimental SOS image generated from more than two steps of raw images is generally unacceptable for its SNR is at least two orders of magnitude lower than a conventional image. Therefore, it is very important to accommodate both resolution and SNR in SOS imaging. 
Here we show that the SNR of a SOS microscopy can be calculated analytically, which can be used as a guideline for researchers to optimize the resolution and SNR performance in SOS imaging.
For an $M$-step SOS, $F_{M\textnormal{-}SOS}(x)=\sum_{i=1}^{M}c_iF_i(x)$, we consider a single pixel at $x=x_0$ and denote the experimentally obtained values of $F_i$ and $F_{M\textnormal{-}SOS}$ as random variables $X_i$ and $Y_{M\textnormal{-}SOS}$, respectively. The mean of $X_i$ is $\mu_i=E[X_i]=\delta(x-x_0)\ast F_i(x)=F_i(x_0)$, which can be calculated using Eq. \ref{eq:steady_state_F} if the excitation intensities and optical parameters are known.
Due to the high detection efficiency of photomultiplier tubes (PMTs) in our confocal and two-photon microscopes, it is valid to consider shot noise as the only noise source in modeling the system. Therefore, the variance of $X_i$, which equals to $\mu_i$, is $\sigma^2_i=\mathrm{var}(X_i)=\mu_i$. 
Since $X_i$ are mutually independent, the mean and variance of an SOS image are $\mu_{M\textnormal{-}SOS}=E[\sum_{i=1}^{M}c_i X_i]=\sum_{i=1}^{M}c_i \mu_i$, $\sigma^2_{M\textnormal{-}SOS}=\mathrm{var}(\sum_{i=1}^{M}c_i X_i)=\sum_{i=1}^{M}c_i^2 \sigma^2_i= \sum_{i=1}^{M}c_i^2 \mu_i$, respectively.
Consequently, the SNR of measuring $Y_{M\textnormal{-}SOS}$ is $SNR_{M\textnormal{-}SOS}={\mu_{M\textnormal{-}SOS}}/{\sigma_{M\textnormal{-}SOS}}={\sum_{i=1}^{M}c_i \mu_i}/{\sqrt{\sum_{i=1}^{M}c_i^2 \mu_i}}$.

Although SOS shares a similar physical principle with SAX microscopy, the implementation of SOS is not only simpler but also superior for its "weak saturation" and versatility. 
SOS works with "weak saturation" where the excitation power is relatively low (e.g., $\sim 3$ mW in 2PEF SOS) compared with the reported excitation intensities used in SAX microscopy (e.g., $\sim 25$ mW in 2PEF SAX) \cite{Nguyen2015}; it permits super-resolution imaging deep in scattering samples within the limit of multiphoton depth when combined with multiphoton microscopy; it also enables super-resolution 3D imaging when combined with depth resolved modalities.
In principle, as long as a conventional fluorescence image with an acceptable SNR can be obtained, regardless of the fluorophores used, the excitation scheme utilized, and the image depth, a corresponding super-resolution SOS image can always be generated.

Potential limitations of SOS microscopy include photobleaching and image mismatch. 
Photobleaching is usually accompanied by high excitation intensity and illumination dose \cite{Bernas2005}. Although SOS microscopy works in the ``weak saturation'' region with low excitation intensities, the prolonged integration time aiming for better SNR performance may cause photobleaching, which degrades imaging quality and resolution. This problem can be mitigated using anti-bleaching agents \cite{Vogelsang2008}, or special fluorophores like quantum dots \cite{Larson2003} and nanodiamonds \cite{Rittweger2009}.
Image mismatch is an issue arising in methods that require subtraction of two images \cite{Kuang2013}. Since SOS microscopy requires the linear combination of several raw images of the same field of view, image mismatch could happen due to sample drift, which can be eliminated using image registration algorithms \cite{Alexander2016}.

\section{Conclusion}
\label{sec:conclusion}
In this paper, we have proposed and demonstrated a new kind of super-resolution fluorescence microscopy using the principle of SOS. 
SOS microscopy employs the ``weak saturation'' phenomenon, which does not require high excitation intensities and therefore the damage to sample can be minimized.
With a set of properly chosen linear combination coefficients, $M$ steps of raw fluorescence images can be linearly combined to generate an $M$-step SOS image, which provides a $\sqrt{M}$-fold increase in spatial resolution compared with conventional diffraction-limited images.
For example, linearly combining two fluorescence images obtained at regular powers extends resolution by a factor of $1.4$ beyond the diffraction limit.
The resolution improvement of an $M$-step SOS microscopy has been validated with simulations.
With the imaging experiments performed on various samples including fixed cells, we have demonstrated the super-resolution capability of the two-step SOS microscopy with both confocal (1PEF) and multiphoton (2PEF) modalities.
Owing to the deep penetration of 2PEF, the multiphoton two-step SOS has been verified to provide super-resolution imaging deep in scattering samples.
The implementation of SOS microscopy is straightforward and requires neither additional hardware nor complex post-processing that are usually required in super-resolution microscopes. A conventional fluorescence microscope can be easily converted into an SOS microscope with super-resolution imaging capability.

\section*{Funding}
National Science Foundation (NSF) (CBET-1554516, DMR-1309587); Defense Advanced Research Projects Agency (DARPA-14-56-A2P-PA-055); American Heart Association (16SDG31230034); Indiana CTSI (NIH/NCRR UL1TR001108).

\section*{Acknowledgments}
Zhang's research was supported by the Berry Family Foundation Graduate Fellowship of Advanced Diagnostics \& Therapeutics (AD\&T), University of Notre Dame. 
Roeder and Nallathamby's research was supported by a grant from the Walther Cancer Foundation. 
The authors further acknowledge the Notre Dame Integrated Imaging Facility (NDIIF) for the use of the A1R-MP confocal microscopy in NDIIF's Optical Microscopy Core, the Notre Dame Center for Nanoscience and Nanotechnology (NDnano), and the Notre Dame Center for Environmental Science and Technology (CEST).

\bibliographystyle{osajnl}
\bibliography{My_reference}

\end{document}